\documentclass[12pt, epsfig]{article}
\usepackage[dvips]{color}
\usepackage{epsfig}%
\usepackage[active]{srcltx}%
\usepackage{amsmath,amsfonts,amssymb,amsthm,amstext,amscd,eucal,srcltx}
\usepackage{epsfig,graphicx,bm}
\usepackage{epstopdf, epsf}
\usepackage{dcolumn}% Align table columns on decimal point
\usepackage{hyperref}
\usepackage{subfigure}
\usepackage{textcomp}
\usepackage{amsfonts}
\usepackage{mathrsfs}
\usepackage{epsfig}
\usepackage{graphicx}%
\usepackage{dcolumn}
\usepackage{amsmath}
\definecolor{darkgreen}{rgb}{0,0.3,0}
\definecolor{darkblue}{rgb}{0,0,0.3}
\definecolor{darkred}{rgb}{0.7,0,0}

\newcommand{\be}{\begin{equation}}
\newcommand{\bse}{\begin{subequations}}
\newcommand{\ese}{\end{subequations}}
\newcommand{\ba}{\begin{array}}
\newcommand{\ea}{\end{array}}
\newcommand{\ee}{\end{equation}}

\makeatletter \@addtoreset{equation}{section}

\def\dre_g{\delta\rho_g}
\def\dpe_g{\delta P_g}
\def\dqe_g{\delta q_g}
\def\dre{\delta\rho}
\def\dpe{\delta P}
\def\dqe{\delta q}

\def\YM1{\frac{\dot\phi^2}{a^2}}
\def\YM2{\frac{g^2\phi^4}{a^4}}

\newcommand{\beq}{\begin{equation}}
\newcommand{\eeq}{\end{equation}}
\newcommand{\bea}{\begin{eqnarray}}
\newcommand{\eea}{\end{eqnarray}}
\newcommand{\bseq}{\begin{subequations}}
\newcommand{\eseq}{\end{subequations}}

\makeatletter
\def\btt#1{\texttt{\@backslashchar#1}}%
\DeclareRobustCommand\bblash{\btt{\@backslashchar}}%
\makeatother
\begin{document}

\title {Cosmology with non-minimal coupled gravity: inflation and perturbation
analysis}
\vspace{3mm}

\author{F. Darabi\thanks{Email: f.darabi@azaruniv.edu}, A. Parsiya\thanks{Email: a.parsiya@azaruniv.edu}\\
{\small Department of Physics, Azarbaijan Shahid Madani University, Tabriz 53714-161, Iran.}}

\maketitle

\begin{abstract}
We study a scalar-tensor cosmological model where the Einstein tensor is non-minimally coupled to the free scalar field dynamics. Using FRW metric,
we investigate the behavior of scale factor for vacuum, matter and dark energy dominated eras. Especially, we focus on the inflationary behavior at early universe. Moreover, we study the perturbation analysis of this model in order to confront the inflation under consideration with the observational results.
\end{abstract}

%\keywords{Inflation,  non-Abelain gauge theory, WMAP data }
%\date{\today}
\maketitle

\section{Introduction}
It is known that the scalar fields play an especially important role in cosmology. For example, one may mention the numerous inflationary models in which inflation is typically driven by a fundamental scalar field so called inflaton. The general form of the action for a scalar-tensor theory with a single scalar field minimally coupled to gravity is given by\footnote{ We have used the units $G=1$.} \cite{linde:2000}
\begin{eqnarray}
S=\int d^4x\sqrt{-g}\left( \frac{R}{16 \pi}+g^{\mu\nu}\nabla_\mu \phi \nabla_\nu \phi- V(\phi)\right),
\end{eqnarray}
where $g_{\mu\nu}$ is the metric tensor, $g=\det(g_{\mu\nu})$, $R$ is the scalar curvature and $V(\phi)$ is the scalar field potential. Also, there are scalar-tensor theories with non-minimal kinetic coupling to the Ricci tensor \cite{Gubitosi:2011}, the scalar curvature $R$ and the f(R) and f(T) theories of modified gravity \cite{Rafael Ferraro:2012}, \cite{Thomas:2005}.

In the application of above gravity theories to the cosmology of early universe, the role of scalar field potential to establish an inflation is unavoidable. However, which potential can exactly describe the correct inflation
at early universe is an open problem. In general, the slowly varying potentials should behave like a large effective cosmological constant suitable for driving an inflation, however the appropriate choice of $V(\phi)$ satisfying the requirements of inflation results in the known problem of fine tuning of the cosmological constant. One way to get rid of the controversial role of the scalar field potential in inflationary models was introduced by Linde as a model so called {\it chaotic inflation} \cite{Linde:1983}
where extremely simple potentials can lead to inflation. Also, some models
have been introduced to represent natural inflation \cite{NI} and inflation with non-minimal derivative coupling \cite{Germani}, \cite{Gra}. One may also work with non-minimal coupling of the scalar field dynamics with Einstein tensor with vanishing or constant  potential to establish the inflationary scenarios   \cite{Ema}, \cite{Gao:2010}. This is a minimal model because it is economic to consider a free scalar field rather than a scalar field subject to a potential term. In other words, it seems more reasonable to think that the universe requires an action to trigger the inflation by the least factors: geometry and free scalar field.

In the present paper, we study an scalar-tensor cosmological model where the Einstein tensor is non-minimally coupled to the free scalar field dynamics.
 We intend to bring this
cosmological model in the framework of a inflation model having suitable slow-roll conditions, an exit mechanism from inflation, creation of baryonic
matter after inflation and so on.
 More importantly, we study the perturbation analysis of this inflation model in order to confront the inflation under consideration with Planck and BICEP2 results. In section II, we study the cosmology with non-minimal kinetic coupled gravity and introduce our inflation model which is followed by matter dominant
and dark energy dominant eras. In section III, we study the cosmic perturbations inside and outside the horizon. In section IV, we study the vacuum fluctuation of the inflaton field to obtain the scalar spectral index and tensor-to-scalar ratio. The paper is ended with a conclusion.

\section{\bf Cosmology with non-minimal kinetic coupled gravity }

 Let us consider a free (without potential term) scalar field whose kinetic term is coupled both with the metric tensor $g_{\mu\nu}$ and Einstein tensor $G_{\mu\nu}$. We write the action as\begin{equation}\label{action}
S=\int d^4x\sqrt{-g}\left[\frac{R}{8 \pi}-( g^{\mu\nu}+\alpha G^{\mu\nu})\nabla_\mu \phi \nabla_\nu \phi-2\Lambda\right]+S_m,
\end{equation}
where $R$ is the Ricci scalar, $\alpha$ is a coupling parameter with dimension of $(length)^2$, $\Lambda$ is a positive
cosmological constant, and $S_m$ is the matter action.
In fact, if one takes the most general action having coupling functions of the metric with the scalar field derivatives, namely
non-minimal coupling of the geometry with the kinetic term of the scalar field,
then it turns out that in order for the field equations to be second order,
one is required to take the action (2.1) \cite{Gubitosi:2011}.
Equations of motion for the scalar field and the metric field are obtained by varying (\ref{action}) with respect to $\phi$ and $g_{\mu\nu}$, respectively
as
\begin{eqnarray}\label{equation of motion}
(g^{\mu\nu}+\alpha G^{\mu\nu})\nabla_\mu \nabla_\nu \phi=0,
\end{eqnarray}
\begin{equation}
G_{\mu\nu}=8\pi [T^{(m)}_{\mu\nu}+T^{(\phi)}_{\mu\nu}+\alpha \Theta_{\mu\nu}],
\end{equation}
where the total energy-momentum tensor is divided into three parts as follows
\begin{equation}
T^{(m)}_{\mu\nu}=(\rho_m+p_m)u_{\mu}u_{\nu}+p_m g_{\mu\nu},
\end{equation}
\begin{equation}
T^{(\phi)}_{\mu\nu}={\nabla_\mu\phi\nabla_\nu\phi-\frac{{1}}{2}g_{\mu\nu}\nabla_\rho\phi\nabla^\rho\phi}-\Lambda
g_{\mu\nu},
\end{equation}
\begin{eqnarray}\label{energy}
 \Theta_{\mu\nu}&=&-{\textstyle\frac12}\nabla_\mu\phi\,\nabla_\nu\phi\,R
+2\nabla_{\alpha}\phi\,\nabla_{(\mu}\phi R^{\alpha}_{\nu)}\nonumber\\
&&
+\nabla^\alpha\phi\,\nabla^\beta\phi\,R_{\mu\alpha\nu\beta}
+\nabla_\mu\nabla^\alpha\phi\,\nabla_\nu\nabla_\alpha\phi
\nonumber\\
&&
-\nabla_\mu\nabla_\nu\phi\,\square\phi\;
-{\textstyle\frac12}(\nabla\phi)^2
G_{\mu\nu}
\nonumber\\
&&
+g_{\mu\nu}\big[-{\textstyle\frac12}\nabla^\alpha\nabla^\beta\phi\,\nabla_\alpha\nabla_\beta\phi
+{\textstyle\frac12}(\square\phi)^2
\nonumber\\
&&
-\nabla_\alpha\phi\,\nabla_\beta\phi\,R^{\alpha\beta}
\big],
\nonumber\\
\end{eqnarray}
where $T^{(m)}_{\mu\nu}$ and $(T^{(\phi)}_{\mu\nu}+\Theta_{\mu\nu})$
are independently conserved.
Using the Friedmann-Robertson-Walker (FRW) flat ($k=0$) background metric with $a(t)$ being the scale factor
\begin{equation}\label{FRW}
{ds^2=-dt^2+a^2(t)(dr^2+r^2d\Omega^2)},
\end{equation}
the field equations are obtained
\beq\label{fieldeq1}
  3H^2=4\pi\dot{\phi}^2\left(1-9\alpha H^2\right) +\Lambda+8\pi\rho_m,
\eeq
\bea\label{fieldeq2}
  2\dot{H}+3H^2&=&-4\pi\dot{\phi}^2
  \left[1+\alpha\left(2\dot{H}+3H^2
+4H\ddot{\phi}\dot{\phi}^{-1}\right)\right]\nonumber \\
&+&\Lambda-8\pi p_m,
\eea
\beq\label{fieldeq3}
  (\ddot\phi+3H\dot\phi)-3\alpha(H^2\ddot\phi
  +2H\dot{H}\dot\phi+3H^3\dot\phi)=0.
\eeq
Therefore, one obtains the total density and pressure respectively as
\begin{equation}\label{rho-T}
{\rho_{_T}=\rho_m+\frac{\Lambda}{8\pi}+\frac{1}{2}\dot{\phi}^2(1-9\alpha H^2)},
 \end{equation}
\begin{equation}\label{p-T}
{p_{_T}=p_m-\frac{\Lambda}{8\pi}+\frac{1}{2}\dot{\phi}^2\left[1+\alpha\left(2\dot{H}+3H^2
+4H\ddot{\phi}\dot{\phi}^{-1}\right)\right],}
\end{equation}
where a dot denotes derivative with respect to $t$. Equation \eqref{fieldeq3}
can be easily integrated to
\begin{equation}\label{phidot}
{\dot{\phi}=\frac{\sqrt{2\lambda}}{a^3(1-3\alpha H^2)}},
\end{equation}
where $\lambda$ is a positive  constant of integration.

In the three following subsections, we will set up a cosmological model in a systematic way which includes an inflation era with suitable slow-roll conditions, an exit mechanism
from inflation, a deceleration era, and an acceleration era of the universe.

\subsection{Inflationary universe}

 By differentiating \eqref{phidot}
with respect to time and substituting $\ddot{\phi}$ and $\dot{\phi}$ in \eqref{rho-T} and \eqref{p-T}, we obtain
\begin{equation}\label{rho-T'}
{\rho_{_T}=\rho_m+\frac{\Lambda}{8\pi}+\frac{\lambda}{a^6(1-3\alpha H^2)^2}(1-9\alpha H^2)},
 \end{equation}
\begin{eqnarray}\label{p-T'}
p_{_T}&=&p_m-\frac{\Lambda}{8\pi}\\ \nonumber
&+&
\frac{\lambda}{a^6(1-3\alpha H^2)^2}\left[1+\alpha\left(2\dot{H}+3H^2
+4H\ddot{\phi}\dot{\phi}^{-1}\right)\right].
\end{eqnarray}

At very early universe where there is no baryonic matter, namely $\rho_m=p_m=0$,
we define $\rho_v=\rho_{_T}$ and $p_v=p_{_T}$ as the density and pressure
of the vacuum state. Then, we combine Eqs.\eqref{rho-T'},
\eqref{p-T'} to obtain the following equation of state
\begin{equation}\label{ES}
p_v=-\rho_v+\frac{2\lambda}{a^6(1-3\alpha H^2)^2}\left[1-9\alpha H^2+\alpha
\dot{H}+\frac{12\alpha^2 H^2\dot{H}}{1-3\alpha H^2}\right].
\end{equation}
Although not all inflationary
models require an equation of state $p_v=-\rho_v$, and even most of them
require an equation of state near it, here we regularly assume  that the inflation model require the vacuum equation of state as $p_v=-\rho_v$. So, in order to achieve this necessary condition for inflation, according to \eqref{ES}, we need to satisfy the following differential equation  for $\lambda\neq 0$
\begin{equation}\label{VES}
\left[(1-9\alpha H^2)+\alpha\dot{H}
\left(1+\frac{12\alpha H^2}{1-3\alpha H^2}\right)\right]=0.
\end{equation}
One may solve this differential equation to obtain the plot $H(t)$
as depicted in Fig.1.
\begin{figure}
\includegraphics[width=9 cm]{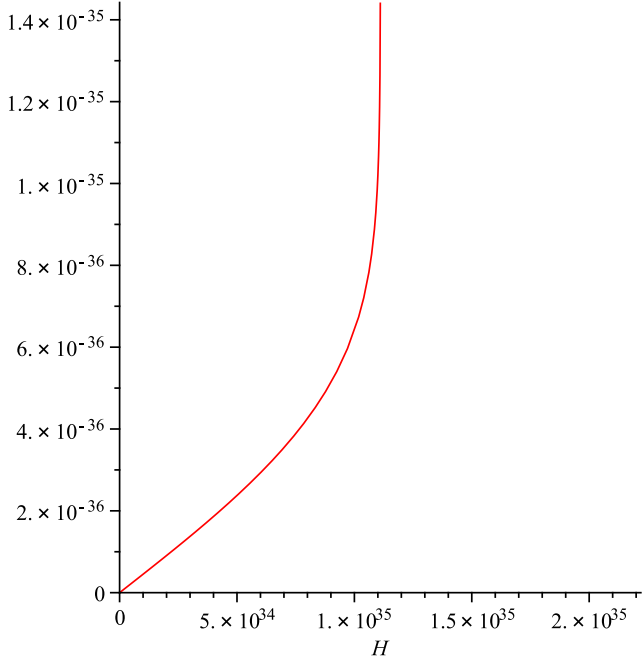}
\caption{\label{fig:deltaphi00} The plot of $H(t)$ for a typical value of
$\alpha=10^{-71} sec^2$ showing the  behaviour $H\simeq1.2 \times10^{35}sec^{-2}$ and ${\dot H}\simeq0$ for $t>10^{-35}$ sec. }
\end{figure}
As is seen in Fig.1, considering a typical small value of $\alpha$, the behaviour for $H(t)$ is so favored to model an inflationary
cosmology with an almost constant ${\dot H}\simeq0$ and large $H$ for $t>10^{-35}$ sec.
Actually, the Hubble parameter $H$ may  suddenly (typically within $10^{-35}$ seconds) approach to the large and almost constant asymptotic value
\begin{equation}\label{H}
H\equiv H_{\alpha}\simeq\sqrt{\frac{1}{9\alpha}} \simeq 1.2 \times10^{35}
sec^{-2},
\end{equation}
provided that  $\alpha\sim10^{-71}sec^{2}$, namely $\alpha$ is assumed to be an infinitesimally small, but non-vanishing parameter\footnote{The non-vanishing
requirement of $\alpha,$ which prevents some expressions like
(\ref{H}) or (\ref{e-folding}) (see below) from diverging,  arises from the following argument. If one lets
$\alpha\rightarrow 0$ , then by using (\ref{phidot}) the equation (\ref{ES}) becomes $p_v=-\rho_v+\dot{\phi}^2$. Now, if we assume a large
kinetic energy $\dot{\phi}^2\gg1$ in the present inflation model, instead of a large potential energy $V({\phi})\gg1$ in the standard
inflation models,  then $p_v\neq-\rho_v$ and we lose the vacuum equation of state $p_v=-\rho_v$ which is necessary for our inflation model.
Therefore,  the limit of $\alpha\rightarrow 0$ together with the base assumption of $\dot{\phi}^2\gg1$ will damage the inflation model, hence we require the
non-vanishing assumption of $\alpha$. In fact, if one is interested in the general relativistic limit of the present non-minimal kinetic coupled inflation model, then the limit $\alpha\rightarrow 0$ should be accompanied
by $\dot{\phi}^2\rightarrow 0$ which is nothing but the so called {\it slow-roll} approximation $\dot{\phi}^2\ll V(\phi)$. In this way, using (\ref{rho-T'}) and (\ref{p-T'}), we
obtain the standard expressions $\rho_{_T}=\frac{\Lambda}{8\pi}$ and $p_{_T}=-\frac{\Lambda}{8\pi}$
in a de Sitter space with vacuum equation of state  $p_v=-\rho_v.$.} The subscript ${\alpha}$ denotes the Hubble parameter at the regime where the contribution of kinetic coupled gravity is dominant.
Therefore, we realize that  imposing the vacuum equation of state results in the typically large and almost constant Hubble parameter \eqref{H}.
Now, we aim to establish an inflation model for the behaviors of scale factor and scalar field dynamics. The cosmological evolution of universe at the vacuum dominated state is described by
\begin{equation}
p_v=-\rho_v,
\end{equation}
\begin{equation}\label{H_{kappa}}
a(t)\propto \exp(H_{\alpha}t),
\end{equation}
and (see \eqref{phidot})
\begin{equation}\label{3H_{kappa}}
\dot{\phi}(t)\propto \exp(-3H_{\alpha}t).
\end{equation}
We know the vacuum era $p_v=-\rho_v$ corresponds to the following approximate
equations
\begin{eqnarray}\label{Einstein1}
3H^2\simeq\frac{1}{3\alpha},
\end{eqnarray}
\begin{eqnarray}\label{Einstein2}
\dot{H}\simeq 0.
\end{eqnarray}
If we define the Hubble slow-roll parameters
\begin{eqnarray}\label{epsilon-eta-rho-P}
\varepsilon\equiv -\frac{\dot H}{H^2}\,,\qquad \eta\equiv\varepsilon-\frac{\dot\varepsilon}{2H\varepsilon},\
\end{eqnarray}
then, as long as $|\varepsilon|\ll 1,\ |\eta|\ll 1$, the inflationary stage is guaranteed.

\subsubsection{Exit mechanism}

In order to study the exit mechanism from inflationary era, we investigate the condition under which the vacuum equation of state
is violated. To this end, we combine \eqref{rho-T'} and \eqref{p-T'} as
\begin{eqnarray}\label{rho-T'-p-T'}
&&p_{_T}+\rho_{_T}\\ \nonumber
&&=\frac{2\lambda}{a^6(1-3\alpha H^2)^2}\left[(1-9\alpha H^2)+\alpha
\dot{H}+\frac{12 \alpha^2 H^2 \dot{H}}{1-3\alpha H^2}\right],
\end{eqnarray}
where, contrary to \eqref{VES}, it is assumed that the RHS will be no longer vanishing when the inflation
reaches to its end, namely $p_v\neq-\rho_v$. On the other hand, combining \eqref{fieldeq1} and \eqref{fieldeq2},
and using \eqref{rho-T'-p-T'} results in
\begin{equation}\label{dot{H}}
-\dot{H}=\frac{8\pi \lambda(1-9\alpha H^2)}{a^6(1-3\alpha H^2)^2+8\pi \lambda
\alpha(1-9\alpha H^2)}.
\end{equation}
Since   during the inflation the asymptotic (maximum) value of  $H$ is ${1}/{3\sqrt{\alpha}}$
(see \eqref{H}),   when we approach to the end of inflation it is strongly
expected that the value of $H$ is bellow this upper bound, so $(1-9\alpha H^2)$ is positive. Moreover, $\lambda>0, \alpha>0$. Therefore, one may expect that when the inflation reaches to its end, according to \eqref{dot{H}}, the quantity $\dot{H}$ is expected to become negative and so $H$ starts decreasing.

On the other hand, substituting \eqref{phidot} in \eqref{fieldeq1}, with $\rho_m=0$ \footnote{Note that $\rho_m$ should become nonzero just after the end of inflation, hence before the inflation is ended, namely when the inflation
is going to its end, we can ignore the matter density.}, we obtain
\begin{equation}\label{{HH}}
{H^2}=\frac{8\pi \lambda(1-9\alpha H^2)}{3a^6(1-3\alpha H^2)^2}+\frac{8\pi}{3} \Lambda.
\end{equation}
Now, by equating \eqref{dot{H}} and \eqref{{HH}} we can find the condition
under which $\varepsilon \rightarrow1$. This results in the following equation
\begin{equation}\label{{slow-roll}}
\frac{2\lambda(1-9\alpha H^2)}{3a^6(1-3\alpha H^2)^2}=\Lambda,
\end{equation}
which is obtained by ignoring the second term in the denominator of \eqref{dot{H}}
due to the small value of the coupling $\alpha$.
Eq.\eqref{{slow-roll}} is the condition under which the slow-roll approximations
$|\varepsilon|\ll 1,\ |\eta|\ll 1$ are violated and the inflation reaches to its end. Therefore, using \eqref{{slow-roll}} we realize that the inflation is ended at $t_f$ when the scale factor is inflated to the order of magnitude
\begin{equation}\label{slow-roll-1}
a(t_f)=\left[\frac{2\lambda(1-9\alpha H^2)}{3\Lambda(1-3\alpha H^2)^2}\right]^{-6}_.
\end{equation}
Note that, although according to \eqref{H_{kappa}} the inflationary expansion of the scale factor is determined just by the coupling parameter $\alpha$,
however, according to \eqref{slow-roll-1} for a given value of $\lambda$ the end of inflation, namely $t_f$, is determined essentially by the cosmological constant $\Lambda$.  Actually, this is a reasonable result because it tells us that both the essential parameters $\alpha$ and $\Lambda$ should be involved in the beginning and ending of the inflation. On the other hand, comparing
\eqref{phidot} with \eqref{slow-roll-1} we find
\begin{equation}\label{Lambda1}
{\dot{\phi}^2(t_f)=\frac{{3\Lambda}}{(1-9\alpha H^2)}}.
\end{equation}
It is seen that, similar to the scale factor, the kinetic energy of scalar field at the end
of inflation is also determined by both of the parameters $\alpha$ and $\Lambda$.
Although, according to \eqref{3H_{kappa}}, the kinetic energy of scalar field
is decreasing with time during inflation, however, from \eqref{Lambda1} it
turns out that for a given value of positive cosmological constant and considering
the very small positive value of $(1-9\alpha H^2)$ one can obtain a large kinetic energy for the scalar field, even at the end of inflation, which can play the
role of a large energy source for reheating the universe after inflation.

\subsubsection{e-folding}

One may obtain the number of e-folding during the inflation as
\begin{equation}\label{e-folding}
N=\int_{t_i}^{t_f} H dt=\int_{t_i}^{t_f} \sqrt{\frac{1}{9\alpha}}dt=\sqrt{\frac{1}{9\alpha}}{(t_f-t_i)}.
\end{equation}
Within the typical short period of time $(t_i=10^{-35})<t<(t_f=10^{-33}),$
required by particle physics, and using $\sqrt{\frac{1}{9\alpha}} \simeq 1.2 \times10^{35}$, we obtain $N\sim120$  well above $N\sim60$ which is at
least needed to overcome the problems of standard cosmology. If we assume that the initial size of the universe before
inflation was about the Planck length $10^{-34} m$, then the 120 number of e-folding results in the final size of the universe at the end of inflation as large as $a(t_f)\sim 10^{22} m$. This large size will remove all the problems
of standard cosmology.

The last important point is to investigate about ghost instability. From
quantum field theory we know that a ghost is a degree of freedom whose propagator has the wrong sign giving rise to a negative norm state on quantisation. A gravity theory with fourth order derivatives in the kinetic term inevitably has ghosts \cite{ghost}.
However, in the case of the kinetic term coupled to the metric and Einstein tensor, the equations of motion for the scalar field is reduced to second
order. Therefore, from physical point of view this theory
can be interpreted as a "good" theory \cite{Felice}. To prevent ghosts in
our model  we insist that the {\it effective kinetic energy} of the scalar field is non-negative. According to Eq.(\ref{fieldeq1}), the effective kinetic
energy of the scalar field is $4\pi\dot{\phi}^2\left(1-9\alpha H^2\right)$. Since $\left(1-9\alpha H^2\right)>0$, the effective kinetic
energy of the scalar field is positive and so the presence of ghosts will be prevented.
\subsection{Radiation and Matter dominated universe}

After the inflationary era the scale factor $a$ becomes exponentially
large (see \eqref{H_{kappa}}), hence the last terms of \eqref{rho-T'}, \eqref{p-T'} containing $a^{-6}$ become ignorable. At this time, namely $t_f$, the very fast decrease in the kinetic energy of the scalar field (see \eqref{3H_{kappa}}) starts and can be balanced by the creation of baryoinc matter with the density and pressure related by the equation of state $p_m=\omega_m\rho_m$. At this stage, we have just two components left as follows
\begin{equation}\label{rho-T1}
\rho_{_T}=\rho_m+\frac{\Lambda}{8\pi},
 \end{equation}
\begin{equation}\label{p-T1}
p_{_T}=p_m-\frac{\Lambda}{8\pi}.
\end{equation}
Assuming an small cosmological constant in comparison with the sufficiently
large values of matter density and pressure, the cosmological evolution of universe at this stage with ignorable cosmological constant is well known as follows \cite{Andrew R. Liddle:2000, Mukhanov:2005}
\\
\\
I)  for $\omega_m=\frac{1}{3}$ we have the radiation dominant era with
the scaling behaviour $\rho_m\propto a^{-4}$ and time evolution
$$a(t) \propto t^{1/2},$$
\\
II)  for $\omega_m=0$ we have the matter dominant era with
the scaling behaviour $\rho_m\propto a^{-3}$  and time evolution
$$a(t) \propto t^{2/3}.$$

\subsection{Dark energy dominated universe}

At the late time and old universe, the scale factor becomes so large that
$\rho_m\propto a^{-3}\ll \Lambda/8\pi$. This stage of evolution is governed
by the cosmological constant
\begin{equation}\label{rho-T2}
\rho_{_T}=\frac{\Lambda}{8\pi},
 \end{equation}
\begin{equation}\label{p-T2}
p_{_T}=-\frac{\Lambda}{8\pi},
\end{equation}
which represents the new phase of vacuum state as
\begin{equation}
p_{_T}=-\rho_{_T}=-\frac{\Lambda}{8\pi},
\end{equation}
where the cosmological constant plays the role of {\it Dark energy}.
The cosmological evolution of this dark energy dominated universe is well known as de Sitter expansion \cite{Andrew R. Liddle:2000, Mukhanov:2005}
\begin{equation}
a(t)\propto \exp(H_{\Lambda}t),
\end{equation}
where
\begin{equation}
H_{\Lambda}=\sqrt{\frac{\Lambda}{3}}.
\end{equation}

\section{\bf  Cosmic perturbation  }

\subsection{\textbf{Background Universe}}

In this section, we focus on the inflationary universe and study the cosmological
perturbations. Let us consider a flat FRW background universe with the metric
\begin{equation}
{ds^2=a^2(\eta)(-d\eta^2+d^2x+d^2y+d^2z))},
 \end{equation}
where $\eta$ is the conformal time defined as $d\eta=dt/a(t)$. In the background universe the scalar field is homogeneous, namely
\begin{equation}
{\bar{\phi}=\bar{\phi}(\eta)}.
 \end{equation}
The background scalar field equation in terms of the conformal time $\eta$ becomes
  \begin{eqnarray}
(-1 &+&3\alpha \frac{\tilde{H}^2}{a^2})\bar{\phi}^{''}-\tilde{H}\acute{\phi}(-1+3\alpha \frac{\tilde{H}^2}{a^2})
\nonumber\\
&+&3\tilde{H}\acute{\bar{{\phi}}}\left( -1+\frac{\alpha}{a^2}(2\acute{\tilde{H}}+\tilde{H}^2)\right) =0,
\;
\end{eqnarray}
where $\tilde{H}=\frac{a^{'}}{a}$  is the Hubble function with respect to the conformal time in the background universe, and $'$ denotes a derivative with respect to the conformal time. Also, the background energy-momentum tensor is given by the following components
\begin{eqnarray}
\bar{T{^0_0}}&=&-\bar{\rho}=-\frac{1}{2}a^{-2}
(-1+9\alpha\frac{\tilde{H}^2}{a^2})\dot{\bar{\phi}}^2,
\nonumber\\
\bar{{T{^0_i}}}&=&0,\
\nonumber\\
\bar{T{^i_j}}&=&\bar{p}\delta{^i_j}\nonumber\\
&=&\dfrac{a^{-2}}{2}\left(    \left( -1
-\frac{\alpha}{a^2}(2\acute{\tilde{H}}-3\tilde{H}^2)\right) \acute{\bar{\phi}}^2
-4\alpha\frac{\tilde{H}\acute{\bar{\phi}}\bar{\phi}^{''}}{a^2}\right).   \nonumber\\
\end{eqnarray}
The Friedman equations are
\begin{eqnarray}
{3\tilde{H}^2}=-8\pi\left(\frac{1}{2}
 (-1+9\alpha\frac{\tilde{H}^2}{a^2})\acute{\bar{\phi}}^2\right),
\end{eqnarray}
\begin{eqnarray}
2\acute{\tilde{H}}&+&\tilde{H}^2=
\\ \nonumber
&-&8\pi\left(\frac{1}{2}(-1-\frac{\alpha}{a^2}(2\acute{\tilde{H}}-3\tilde{H}^2))\acute{\bar{\phi}}^2-2\alpha\frac{\tilde{H}\acute{\bar{\phi}}\bar{\phi}^{''}}{a^2}\right),
\end{eqnarray}
\begin{eqnarray}
\acute{\tilde{H}}=-\frac{8\pi}{3}\left((-1
 -\frac{3\alpha}{2a^2}(\acute{\tilde{H}}-3\tilde{H}^2))\acute{\bar{\phi}}^2
 -3\alpha \tilde{H}\acute{\bar{\phi}}\bar{\phi}^{''}\right).\nonumber\\
\end{eqnarray}

\subsection{\textbf{ Perturbed Universe  in the Newtonian Gauge}}
The metric of perturbed universe in the Newtonian gauge is
\begin{equation}
{g_{\mu\nu}=\bar{g}_{\mu\nu}+\delta g_{\mu\nu}},
\end{equation}
and
\begin{equation}\label{perturb metric}
{ds^2=a^2(\eta)\left( -(1+2\Psi)d^2\eta+(1-2\Psi)\delta_{ij}dx^idx^j\right), }
\end{equation}
where $\Psi$ is the gauge invariant Newtonian potential which characterizes the metric perturbations. The determinant of the perturbed metric is obtained
\begin{equation}
{g=-a^8(1-4\Psi).}
\end{equation}
In this perturbed metric, the scalar field equation takes the form
\begin{eqnarray}\label{perturbation}
 &&g{^0_0}(1+\alpha G{^0_0})(\phi^{''}-\Gamma_{00}^0\phi^{'}-\Gamma_{00}^i\phi_{,i})
 \nonumber\\
 &+&g{^i_i}(1+\alpha G{^i_i})(\nabla^2{\phi}-\Gamma_{ii}^0\phi^{'}-\Gamma_{ii}^j\phi_{,j})
\nonumber\\
&+&\frac{2\alpha}{a^2}G^0_i(\phi_{,i}^{'}
-\Gamma_{0i}^j\phi_{,j})=0.
\end{eqnarray}
Now, we divide the scalar field into a background and a perturbed parts
\begin{equation}\label{phi perturbe}
{\phi=\bar{\phi}(\eta)+\delta\phi(\eta,\vec{x})}.
\end{equation}
Also, the required perturbed geometrical quantities in the Newtonian gauge are \cite{Kurki-Suonio:2011},
\begin{eqnarray}\label{general perturbed}
 \Gamma_{00}^0&=&\tilde{H}+\acute{\Psi}\,,
 \nonumber\\
  \Gamma_{0i}^0&=&\acute{\Psi}_{,i}\,,
  \nonumber\\
   \Gamma_{ij}^0&=&\tilde{H}(1-4\acute{\Psi})
   \delta_{ij}-\delta_{ij}\acute{\Psi}\,,
   \nonumber\\
  \Gamma_{00}^i&=&-\acute{\Psi}_{,i}\,,
  \nonumber\\
 \Gamma_{0j}^i&=&\tilde{H}\delta_{ij}-\delta_{ij}\acute{\Psi}\,,
 \nonumber\\
 \Gamma_{jk}^i&=&-\delta_j^i\Psi_{,k}-\delta_k^i\Psi_{,j}+\delta_{jk}\Psi_{,i}\,,
 \nonumber\\
 G_0^0&=&-3a^{-2}\tilde{H}^2+a^{-2}\{-2\nabla^2\Psi+6\tilde{H}\acute{\Psi}+6\tilde{H}^2\Psi\}\,,
  \nonumber\\
 G_0^i&=&-G_i^0=a^{-2}\{-2\acute{\Psi}_{,i}-2\tilde{H}\Psi_{,i}\}\,,
 \nonumber\\
 G_i^j&=&a^{-2}(-2\tilde{H}^{'}-\tilde{H}^2)\delta_{ij}\,,
\nonumber\\
 &+&a^{-2}\left( 2\Psi^{''}
 +6\tilde{H}\acute{\Psi}+(4\tilde{H}^{'}
 +2\tilde{H}^2)\Psi\right).
  \end{eqnarray}
Substituting Eqs.\eqref{general perturbed} and \eqref{phi perturbe} into \eqref{perturbation}, we get the field perturbation equation as follows
\begin{eqnarray}\label{field perturbation equation}
 (-1+\frac{3\alpha}{a^2}\tilde{H}^2)(\delta\phi)^{''}
 +2\tilde{H}(-1+\frac{3\alpha}{a^2}\tilde{H}^{'})(\delta\phi)^{'}
 \nonumber\\
 -(-1+\frac{\alpha}{a^2}(2\tilde{H}^{'}+\tilde{H}^2))\nabla^2(\delta\phi)=-Ia^2,
  \end{eqnarray}
where
\begin{eqnarray}\label{I}
I&=&\frac{1}{a^2}(-1+\frac{3\alpha\textsl{}}{a^2})(-\Psi^{'}\bar\phi^{'})
\nonumber\\
&+&\frac{1}{a^{2}}\{ -\frac{\alpha}{a^2}
(-2\nabla^2\Psi+6\tilde{H}\Psi^{'}+6\tilde{H^2}\Psi)
 \nonumber\\
 &+&2\Psi-\frac{6\alpha\Psi}{a^2}\tilde{H}^2\}(\bar\phi^{''}-\tilde{H}\bar\phi^{'})
\nonumber\\
 &-&\dfrac{3\tilde{H}}{a^2}\lbrace \dfrac{\alpha}{a^2}\{+2\Psi^{''}
+6\tilde{H}\Psi^{'}+(4\tilde{H}^{'}+2\tilde{H}^2)\Psi\}
\nonumber\\
 &+&2\Psi+\dfrac{2\alpha\Psi}{a^2}(-2\tilde{H}^{'}-\tilde{H}^2)\rbrace\bar\phi^{'}
\nonumber\\
 &-&\frac{3(4\Psi\tilde{H}+\Psi^{'})}{a^2}
 \left( -1+\frac{\alpha}{a^2}(2\tilde{H}^{'}+\tilde{H}^2) \right) \bar\phi^{'}.
  \end{eqnarray}
Using the slow-roll approximation $\dot H\approx0$ and Eqs.\eqref{field perturbation equation} and \eqref{I} become
\begin{eqnarray}\label{field perturbation equation 2}
(-1+ \frac{3\alpha}{a^2}\tilde{H}^2)\{(\delta\phi)^{''}
 +2\tilde{H}(\delta\phi)^{'}&-&\nabla^2(\delta\phi)\}
 \nonumber\\
&=&-I^{'}a^2,
  \end{eqnarray}
where
  \begin{eqnarray}\label{II}
 I^{'}&=&\frac{1}{a^2}(-1+\frac{3\alpha\textsl{}}{a^2})(-\Psi^{'}\bar\phi^{'})
 \nonumber\\
 &+&\frac{1}{a^{2}}(-\frac{\alpha}{a^2}
(-2\nabla^2\Psi+6\tilde{H}\Psi^{'}+6\tilde{H^2}\Psi)+2 \Psi
 \nonumber\\
 &-&\frac{6\alpha\Psi}{a^2}\tilde{H}^2)(\bar\phi^{''}-\tilde{H}\bar\phi^{'})
\nonumber\\
 &-&\dfrac{3\tilde{H}}{a^2}\lbrace\frac{\alpha\textsl{}}{a^2} (2\Psi^{''}+6\tilde{H}\Psi^{'}+6\tilde{H}^2\Psi)
\nonumber\\
 &+&2\Psi-\frac{6\alpha\textsl{}\Psi}{a^2}\tilde{H}^2\rbrace\bar\phi^{'}
\nonumber\\
&-&\frac{3(4\Psi\tilde{H}+\Psi^{'})}{a^2}
 \left(-1+ \frac{3\alpha}{a^2}\tilde{H}^2 \right) \bar\phi^{'}.
  \end{eqnarray}
In order to obtain the equations of perturbations, we have to linearize the Einstein equations
\begin{equation}
{G_\nu^\mu=R_\nu^\mu-\frac{1}{2}\delta_\nu^\mu R=8\pi T_\nu^\mu},
 \end{equation}
for small inhomogeneities about FRW background universe. The components
of Einstein tensor for the background metric \eqref{FRW} are obtained easily and result in \cite{Mukhanov:2005}
\begin{eqnarray}
\bar{G}_0^0=\frac{3\tilde{H}^2}{a^2},\bar{G}_i^0=0,
\end{eqnarray}
\begin{eqnarray}
\bar{G}_i^j=\frac{1}{a^2}(2\acute{\tilde{H}}+\tilde{H}^2)\delta_{ij}.
\end{eqnarray}

The linearized equations for perturbations are
\begin{equation}\label{deltaG}
{\delta G_\nu^\mu=8\pi\delta T_\nu^\mu }.
 \end{equation}
The direct calculation of \eqref{deltaG} for the metric \eqref{perturb metric} gives the equations \cite{Mukhanov:2005}
\begin{equation}\label{22'}
{\nabla^2\Psi-3\tilde{H}(\acute{\Psi}+\tilde{H}\Psi)=4\pi a^2\delta T_0^0 },
 \end{equation}
\begin{equation}\label{22}
{(\acute{\Psi}+\tilde{H}\Psi)_{,i}=4\pi a^2\delta T_i^0 },
 \end{equation}
 \begin{equation}\label{22''}
{\left( \Psi^{''}+3\tilde{H}\acute{\Psi}+(2\acute{\tilde{H}}+\tilde{H}^2)\Psi\right) \delta_ij=-4\pi a^2\delta T_j^i }.
 \end{equation}
Equation \eqref{field perturbation equation} contains two unknown variables, $\delta\phi$ and $\Psi$, and should be supplemented by one of the equations
\eqref{22'}-\eqref{22''}. It is convenient to use \eqref{22}. Substituting $\phi=\bar{\phi}+\delta\phi(\eta,\vec{x})$ in \eqref{energy}, using Eqs.\eqref{general perturbed}, and after a tedious but straightforward calculation we obtain
\begin{eqnarray}
\delta T_i^0
&=&a^{-2}\left( 1+6\alpha a^{-2}(2\tilde{H}^{'}+\tilde{H}^2)\right)
\acute{\bar{\phi}}\delta\phi_{,i}
\nonumber\\
&+&4\alpha a^{-4}\Psi^{'}_{,i}\acute{\bar{\phi}}^2.
\end{eqnarray}
Hence, equation \eqref{22} becomes
\begin{eqnarray}\label{2}
 \acute{\Psi}+\tilde{H}\Psi
&=&4\pi\lbrace (1+{6\alpha}{a^{-2}}(2\tilde{H}^{'}+\tilde{H}^2))\acute{\bar{\phi}}\delta\phi\
\nonumber\\
&+&4\alpha a^{-4}\Psi^{'}\acute{\bar{\phi}}^2\rbrace.
  \end{eqnarray}
Now, we solve \eqref{field perturbation equation} and \eqref{2} in two limiting cases: i) for short wavelength perturbations where the physical wavelength $\lambda_{ph}$ is much smaller than the curvature scale $H^{-1}$, ii) for long wavelength perturbations where $\lambda_{ph}$ is much larger than the curvature scale $H^{-1}$. Since the curvature scale does not change very much during the inflation and the physical scale of perturbations grows like $\lambda\sim a/k $, hence we are interested in the short-wavelength perturbation where the physical wavelength starts smaller than the Hubble length but eventually exceeds it. We may fix the amplitude of these modes by vacuum fluctuations
through the uncertainty principle. Then, we study how the amplitude of these perturbations evolves after it crosses the Hubble length.

\vspace{5mm}

%%%%%%%%%%%%%%%%%%%%%%%%%%%%%%%%%%%%%%%%%%%%%%%%%%%%%%%%%%%%%%%%%%%%%%%%%%%%%%%%%%%%%%%%%%%
\subsection{\bf  Inside the Hubble scale }

For the short-wavelength perturbations we have $\lambda_{ph}\ll H^{-1}$ or, equivalently $k\gg Ha\sim|\eta|^{-1}$. Moreover, for a very large $k \eta$ the spatial derivative term dominates in \eqref{field perturbation equation} and its solution becomes as $\exp(\pm ik\eta)$ to leading order. On the other hand, the gravitational field oscillates ($\acute{\Psi}\sim k\Psi$), and can be estimated from \eqref{2} as
\begin{equation}\label{varphi}
 {\Psi=\frac{4\pi}{k}\left( \frac{1+6\alpha a^{-2}(2\tilde{H}^{'}+\tilde{H}^2)}{1-4\alpha a^{-2}\acute{\bar{\phi}}^2}\right) \bar{\phi}^{'}\delta\phi,}
\end{equation}
where $\acute{\bar\phi}=a\dot{\bar\phi}$. Taking into account that during the inflation $\dot{H}\approx0$, we will get
\begin{equation}
 {\Psi=\frac{4\pi}{k}\left( \frac{1+18\alpha a^{-2}\tilde{H}^2}{1-4\alpha a^{-2}\acute{\bar{\phi}}^2}\right) \bar{\phi}^{'}\delta\phi}.
\end{equation}
Now, using this equation  and also $k\gg Ha$, after a tedious but straightforward calculation we find that the equation \eqref{field perturbation equation 2} reduces to the following equation in terms of physical time
\begin{eqnarray}
\ddot{({\delta{\phi}})} +3{H}\dot{({\delta{\phi}})}+\frac{k^2}{a^2}(\delta\phi)
 \approx 0,
  \end{eqnarray}
where use has been made of the slow-roll approximation. This equation is
easily solved with $H$ being constant. Figure 2, shows the behavior of $\delta\phi$
as a function of the Hubble time $(Ht)$.
\begin{figure}
\includegraphics[width=9 cm]{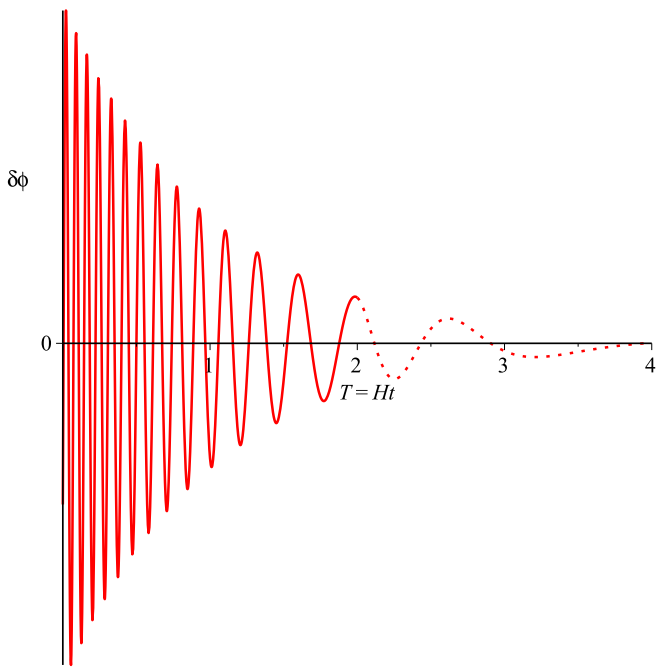}
\caption{\label{fig:deltaphi11}The behavior of short-wavelength perturbations  using the slow-roll approximation $H\simeq\mbox{constant}$, for $k\gg Ha$, and $a_0=1$. The doted line represents that part of perturbation  which is discarded due to crossing the horizon, after a typical value $T\gtrsim2$.}
\end{figure}

\vspace{5mm}
%%%%%%%%%%%%%%%%%%%%%%%%%%%%%%%%%%%%%%%%%%%%%%%%%%%%%%%%%%%%%%%%%%%%%%%%%%%%%%%%%%%%%%
\subsection{\textbf{Evolution through Horizon Exit }}

To take advantage of the slow-roll approximation for the perturbations, we need to recast \eqref{field perturbation equation} and \eqref{2} in terms of the physical time $t$ as
\begin{eqnarray}\label{deltaphi}
(-1 &+&3\alpha {H}^2)\ddot{({\delta{\phi}})}+3H\left({-1} +\alpha(2\dot{H}+3{H}^2)\right) \dot{({\delta{\phi}})}
 \nonumber\\
 &-&\left({-1} +\alpha(2\dot{H}+3{H}^2)\right) \nabla^2(\delta\phi)=-I,
\end{eqnarray}
\begin{eqnarray}
 \dot{\Psi}+H\Psi
 =4\pi\{(-1+6\alpha(2\dot{H}+3H^2))\dot{\bar\phi}\delta\phi+4\alpha\dot{\Psi}\dot{\bar\phi}^2\},
  \end{eqnarray}
where, $I$ is the quantity defined in Eq.\eqref{I} which is expressed here
in terms of the physical time $t$. The spatial derivative term $\nabla^2\phi\approx k^2\phi$ can be neglected for long-wavelength inhomogeneities (i.e. $k\ll Ha$).
To find the non-decaying slow-roll mode we next omit terms proportional to $\dot{\Psi}$ \cite{Mukhanov:2005}. The equations for the perturbations in slow-roll regime result in
\begin{eqnarray}\label{deltaphii}
(-1&+&3\alpha H^2)\{\ddot{({\delta{\phi}})}+3H\dot{({\delta{\phi}})}\} \nonumber\\
&=&-2\Psi(-1+6\alpha H^2)(\ddot{\phi}+3H\dot{\phi}),
\end{eqnarray}
\begin{equation}\label{varphii}
 {\Psi=\frac{4\pi}{H}(1+18\alpha H^2)\dot{\phi}(\delta\phi),}
  \end{equation}
where we have removed the overbears on the background quantities, for simplicity.
But, in the slow-roll approximation the scalar field equation is,
\begin{eqnarray}
 (-1+3\alpha H^2)(\ddot{\phi}+3H\dot{\phi})\approx0.
  \end{eqnarray}

 Then, we can ignore the RHS of Eq.\eqref{deltaphii}, and rewrite \eqref{deltaphii} as
\begin{eqnarray}
 \ddot{({\delta{\phi}})}+3H\dot{({\delta{\phi}})}\approx 0.
   \end{eqnarray}
This equation is also easily solved with $H$ being constant. The solution of $\delta\phi$ can be set to  depict a behavior as shown in Fig.3, where $\delta\phi$ is almost constant for $Ht\gtrsim 2$. 
\begin{figure}
\includegraphics[width=9 cm]{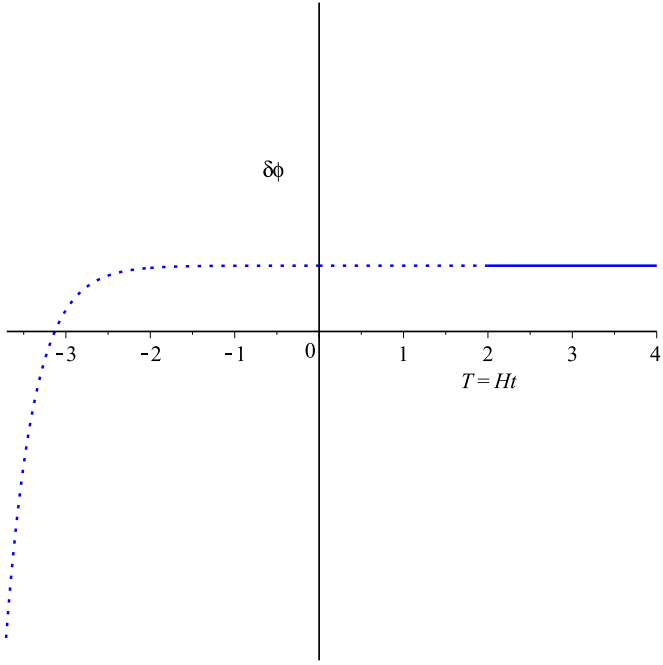}
\caption{\label{fig:deltaphi22} The behavior of long-wavelength perturbations using the slow-roll approximation  $H\simeq\mbox{constant}$, for $k\ll Ha$, and $a_0=1$. It is seen that $\delta\phi$ is almost constant for $T>0$, specifically
 after crossing the horizon at a typical value $T\gtrsim2$, shown by solid
 line.}
\end{figure}

 In conclusion, for $a\lesssim a_k\sim\frac{k}{H}$ the perturbations are inside the horizon and their amplitudes decrease with time, as is seen in Fig.2. After several Hubble time, for example $Ht\gtrsim 2$, the perturbations cross the horizon, for $a\gtrsim a_k$, and their amplitudes freeze out at their last values crossing the horizon, as is seen in Fig.3. Combining two perturbations inside and outside the horizon,
we may obtain the desired behavior of perturbations as is depicted in Fig.4.
\begin{figure}
\includegraphics[width=9 cm]{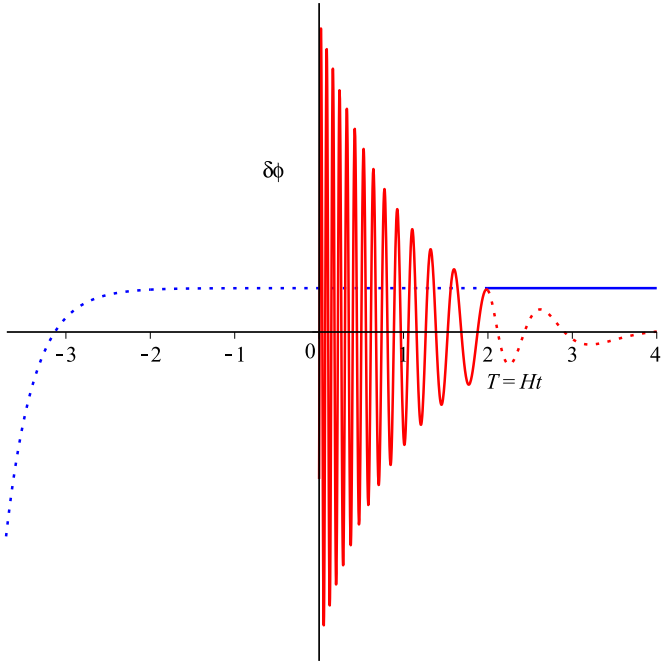}
\caption{\label{fig:deltaphi23} The typical behavior of short and long wavelength
perturbations, inside and outside of the horizon, respectively  for $0<T\lesssim2
$ and $T\gtrsim2$.}
\end{figure}

\vspace{5mm}
%%%%%%%%%%%%%%%%%%%%%%%%%%%%%%%%%%%%%%%%%%%%%%%%%%%%%%%%%%%%%%%%%%%%%%%%%%%%%%%%%%%%%%%%%%%
\section{\textbf{Vacuum fluctuation of the inflaton field  }}

  The scalar field perturbation
equation, in the absence of metric perturbation is obtained as
\begin{equation}\label{phi perturbe2}
(-1+3\alpha H^2)(\delta\ddot{{\phi}}+3H\delta\dot{{\phi}}
-\bigtriangledown^2\delta{\phi})
=0,
\end{equation}
where use has been made of  (\ref{phi perturbe}), (\ref{fieldeq3}) and
 the slow-roll approximation $\dot{H}\approx0,$ in the scalar field equation \eqref{equation of motion}.
Using
 (\ref{Einstein1}) which results in $(-1+3\alpha H^2)\neq 0$,
the equation \eqref{phi perturbe2}  in the conformal time ($\eta$) is rewritten as
\begin{equation}\label{phi perturbe3333}
{{\delta{\phi}}^{''}+2\tilde{H}{\delta{\phi}}^{'}-\bigtriangledown^2\delta{\phi}
=0}.
\end{equation}
For a given Fourier component having wave number $k$, and by defining $u\equiv a \delta{\phi}_k$, we obtain\footnote{Note that the perturbation analysis
of the scalar field equation, described by \eqref{phi perturbe5''}, is almost indistinguishable from that of studied in the usual inflationary models, 
up to terms including slow-roll parameters in the RHS (see Ref.[13]). This agreement may be a good point, because we obtain the same perturbative analysis of common inflation models (including a scalar field potential) in our inflation model (not including a scalar field potential). Of course, if we would consider the metric perturbations too, then we could obtain highly  nonlinear and complicate perturbative differential equations with possibly different perturbative analysis. To avoid this complexity, and just for simplicity, we have  considered the scalar field perturbations over a fixed background.}
\begin{equation}\label{phi perturbe5''}
{{u}^{''}+\{k^2-\frac{1}{\eta^2}(3\varepsilon+2)\}u=0},
\end{equation} 
where use has been made of slow-roll regime to keep terms up to $1^{th}$ order in the slow-roll parameter $\varepsilon$, namely $\varepsilon^2\approx 0$. By defining a new function $S$ so that
 $u\equiv(-\eta)^{\frac{1}{2}}S $, we obtain
\begin{equation}\label{phi perturbe53}
{\eta^2{S}^{''}+\eta S^{'}+(k^2\eta^2-\nu^2)u=0},\:\:\:\:\:\:\:(\nu\approx\frac{3}{2}+\varepsilon)
\end{equation}
whose solutions are the Hankel functions. For early time ($-k\eta\longrightarrow \infty$) and late time ($-k\eta\longrightarrow 0$) we have respectively \cite{Kurki-Suonio:2011}
\begin{eqnarray}\label{deltaphi10}
{\delta{\phi}}_{k}=C_k\sqrt{\frac{2}{\pi}}\frac{1}{a\sqrt{k}}e^{-i k\eta},
\end{eqnarray}
\begin{eqnarray}\label{deltaphi1}
{\delta{\phi}}_{k}=C_k a^{-1}\sqrt{-\eta}\sqrt{\frac{2}{\pi}}2^{\nu-\frac{3}{2}}\frac{\Gamma(\nu)}{\Gamma(\frac{3}{2})}({- k\eta})^{-\nu}\propto(-\eta)^{\frac{3}{2}+\varepsilon -\nu}.
   \end{eqnarray}
Therefore, using $\nu\approx\frac{3}{2}+\varepsilon$, the perturbation ${\delta{\phi}_{k}}$ at late time becomes almost independent of the conformal time.
The power spectrum of these vacuum fluctuations is defined as \cite{Lyth}
\begin{eqnarray}\label{deltaphi111}
P_\phi(k)=L^3\frac{k^3}{2\pi^2}\langle(\delta{\phi})^2\rangle.
   \end{eqnarray}
Well bellow the horizon exit, $k\gg\tilde{H}$,  the field operator ${\delta{\phi}}_k(\eta)$ becomes the Minkowski space field operator and we have standard typical vacuum fluctuations in  $\phi$. Well above the horizon exit, scale dependence of
the power spectrum of $\delta{\phi}$ fluctuations is obtained as \cite{Kurki-Suonio:2011}
\begin{eqnarray}\label{deltaphi11111}
P_\phi(k,\eta)\propto{ k}^{3-2\nu}\propto{ k}^{-2\varepsilon}\equiv k^{n_{s}-1},
   \end{eqnarray}
which leads to the scalar spectral index
\begin{eqnarray}\label{deltaphi111111}
n_{s}=1-2\varepsilon.
   \end{eqnarray}
The tensor-to-scalar ratio is also obtained as \cite{Bruce}
   \begin{eqnarray}\label{deltaphi62}
r=\frac{P_T}{P_R}=\left[\frac{\dot{\phi}^2}{\pi H^2}\right]_{k=aH}=\left[\frac{9{\phi}^2}{\pi }\right]_{k=aH},
   \end{eqnarray}
where use has been made of Eqs.\eqref{3H_{kappa}}, \eqref{e-folding} together
with $\frac{H}{\dot{\phi}}=\frac{dN}{d\phi}$, $\dot{H}\simeq 0$.
Comparing with the upper bounds obtained by the recent Planck and BICEP2 measurements \cite{2-i,2-m,BI}
\begin{equation}\label{ns}
 n_s=0.9603\pm0.0073\:,\:\:\:\:\:r\simeq\left\{
\begin{array}{ll} 0.11\, \qquad  &Planck
\\
0.2\,\,\,\,\,\,\,\,\,\,\,\,\,\, \qquad &BICEP2
\end{array}\right.,
\end{equation}
we find $\varepsilon\simeq 2\times 10^{-2}$ and
\begin{equation}\label{ns1}
\left[{\phi}\right]_{k=aH}\simeq\left\{
\begin{array}{ll} 0.19\,\, M_{P}\, \qquad  &Planck
\\
0.26\,\, M_{P}\,\,\,\,\,\,\,\,\,\,\,\,\,\, \qquad &BICEP2,
\end{array}\right.
\end{equation}
which predicts the vacuum expectation value of the scalar field, at the time
of leaving the horizon, in terms of the Planck mass.
%%%%%%%%%%%%%%%%%%%%%%%%%%%%%%%%%%%%%%%%%%%%%%%%%%%%%%%%%%%%%%%%%%%%%%%%%%%%%%%%%%%%%%%%%%%
\smallskip
\section{Conclusion and discussion}

Motivated by the fact that the common inflationary scenarios usually need a scalar field potential to trigger the inflation, and that taking the proper inflaton potential without fine tuning and cosmological constant problems is still an unsolved issue, we have studied the cosmological implications of a kinetic coupled scalar-tensor gravity to establish a systematic inflation
model which is capable of transition to the matter dominant and dark energy
dominant eras. Moreover, we have studied the perturbation analysis of this inflation model in order to confront the inflation under discussion with the recent observational results. Note that the simplified perturbation analysis followed here, leads to no constraints on the parameter $\alpha$.

\section*{Acknowledgment}
We would like to thank the anonymous referees whose useful comments much improved the presentation of this manuscript. This research has been supported by Azarbaijan Shahid Madani university by a research fund No. 403.16.

\newcommand\AL[3]{~Astron. Lett.{\bf ~#1}, #2~ (#3)}
\newcommand\AP[3]{~Astropart. Phys.{\bf ~#1}, #2~ (#3)}
\newcommand\AJ[3]{~Astron. J.{\bf ~#1}, #2~(#3)}
\newcommand\APJ[3]{~Astrophys. J.{\bf ~#1}, #2~ (#3)}
\newcommand\APJL[3]{~Astrophys. J. Lett. {\bf ~#1}, L#2~(#3)}
\newcommand\APJS[3]{~Astrophys. J. Suppl. Ser.{\bf ~#1}, #2~(#3)}
\newcommand\JHEP[3]{~JHEP.{\bf ~#1}, #2~(#3)}
\newcommand\JCAP[3]{~JCAP. {\bf ~#1}, #2~ (#3)}
\newcommand\LRR[3]{~Living Rev. Relativity. {\bf ~#1}, #2~ (#3)}
\newcommand\MNRAS[3]{~Mon. Not. R. Astron. Soc.{\bf ~#1}, #2~(#3)}
\newcommand\MNRASL[3]{~Mon. Not. R. Astron. Soc.{\bf ~#1}, L#2~(#3)}
\newcommand\NPB[3]{~Nucl. Phys. B{\bf ~#1}, #2~(#3)}
\newcommand\PLB[3]{~Phys. Lett. B{\bf ~#1}, #2~(#3)}
\newcommand\PRL[3]{~Phys. Rev. Lett.{\bf ~#1}, #2~(#3)}
\newcommand\PR[3]{~Phys. Rep.{\bf ~#1}, #2~(#3)}
\newcommand\PRD[3]{~Phys. Rev. D{\bf ~#1}, #2~(#3)}
\newcommand\RMP[3]{~Rev. Mod. Phys.{\bf ~#1}, #2~(#3)}
\newcommand\SJNP[3]{~Sov. J. Nucl. Phys.{\bf ~#1}, #2~(#3)}
\newcommand\ZPC[3]{~Z. Phys. C{\bf ~#1}, #2~(#3)}
\newcommand\IJGMMP[3]{~Int. J. Geom. Meth. Mod. Phys{\bf ~#1}, #2~(#3)}
\newcommand\CTP[3]{~Commun. Theor. Phys.{\bf ~#1}, #2~(#3)}
\newcommand\CQG[3]{~ Class. Quant. Grav.{\bf ~#1}, #2~(#3)}
\newcommand\MPLA[3]{~ Mod. Phys. Lett. A{\bf ~#1}, #2~(#3)}

\end{document}